\documentstyle[aps,psfig,preprint,graphics]{revtex}
\def\k{{\bf k}}
\def\Q{{\bf Q}}
\def\D{{\cal D}}

\begin{document} 
\draft
\date{\today}
\title{Non-perturbative gluons in diffractive photo-production of $J/\Psi$}

\author{M. B. Gay Ducati~\thanks{gay@if.ufrgs.br}, Werner K. Sauter~\thanks{sauter@if.ufrgs.br}}

\address{{\it Departamento de F\'{\i}sica, Universidade Federal do Rio Grande do Sul, Porto Alegre, Brasil }}
\maketitle

\begin{abstract}
The modifications induced in the calculation of the cross section of the diffractive process $\gamma \gamma \rightarrow J/\Psi J/\psi$ when the gluon propagator is changed are analyzed. Instead of the usual perturbative gluon propagator, alternative forms obtained using non-perturbative methods like Dyson-Schwinger equations are used to consider in a more consistent way the contributions of the infrared region. The result shows a reduction in the differential cross-section for low momentum transfer once compared with the perturbative result, to be confirmed with future experimental results from TESLA. 
\end{abstract}

\thispagestyle{empty}  
\newpage
\setcounter{page}{1}

The description of the diffractive processes in high energy physics is still an open question. In the high energy limit, those processes are usually described by a Pomeron exchange~\cite{fr}, whose nature remains one of the most intriguing phenomena in Quantum Chromodynamics (QCD). Its properties were first described by the Regge theory~\cite{coll} and used to explain the features of total hadronic cross-sections and differential cross-sections at small transverse momenta (the soft Pomeron), whose properties relied on  the non-perturbative sector of the QCD. In high energy QCD, the Pomeron corresponds to the sum of gluonic ladder diagrams, with reggeized gluons in the vertical lines, described by the Balitski\v{\i}-Fadin-Kuraev-Lipatov (BFKL) equation (the hard Pomeron)~\cite{bfkl}. Although the BFKL equation is infrared finite, there are difficulties to control contributions from the infrared region, reflecting in the fact that the BFKL approach does not give the same results as the Regge theory for the soft Pomeron. Since perturbative QCD can only be applied when all momenta are sufficiently large~\cite{fr}, in a region of low momentum transfer we can expect an overlap of the effects from the soft and hard sectors.

In an experimental procedure to test the hard Pomeron, the observables should have some particular properties: the virtuality of the gluons in the ladders must be sufficiently large to enable the use of a perturbative expansion and  the hard scale is provided by the coupling of the ladder to the incoming particles or by the momentum of the gluons in the ladder.

\begin{figure}[ht]
\begin{center}
\scalebox{.75}{\includegraphics*[140pt,430pt][520pt,770pt]{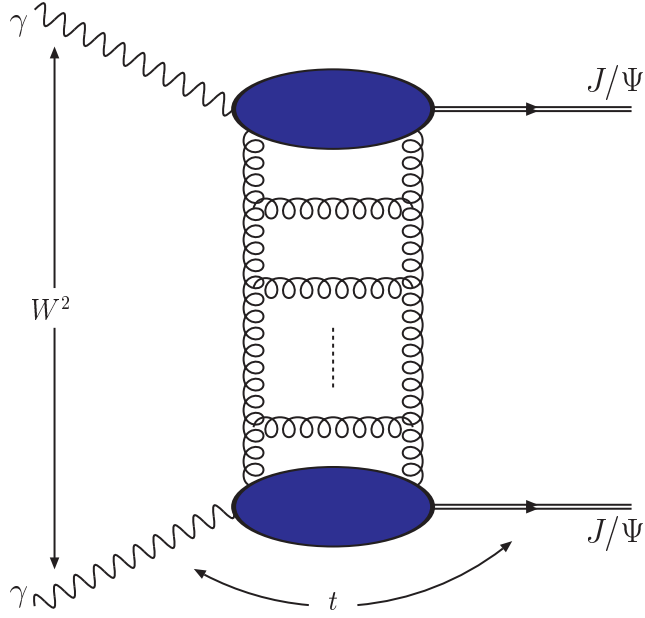}}
\end{center}
\caption{~The process $\gamma \gamma \rightarrow J/\Psi \, J/\Psi$ with the exchange of the QCD Pomeron.}
\label{f1}
\end{figure}

Due to the requirements above, measurements in hadron-hadron and lepton-hadron colliders are limited. Kwieci\'nski and  Motyka~\cite{km98} argued that these difficulties can be  avoided in the case of the production of massive vector mesons from two photon reactions at TESLA~\cite{der01}. The exclusive process $\gamma \gamma \rightarrow J/\Psi\;J/\Psi$ (see fig.[\ref{f1}]) can test the exchange of the Pomeron in arbitrary momentum transfer since the hard scale is given by the (large) mass of the quark charm, for both sides of the diagram, instead of the gluon momenta in the ladder. However, in the infrared region (low momentum transfer) where the soft Pomeron is more important, we expect non-perturbative effects will play a crucial role in the description of these processes. To estimate these soft effects, in this letter we propose as a first step the replacement of the usual perturbative gluon propagator by distinct ones that include non-perturbative effects, considering the soft Pomeron as the exchange of two non-perturbative gluons. This model for the Pomeron was used successfully to describe another soft processes as the elastic proton-proton scattering using the gluon propagator from Dyson-Schwinger equations (DSE)~\cite{hkn93} or from lattice field theory (LFT)~\cite{hpr96} and the elastic electro-production of vector mesons~\cite{gdhn93}. The modified propagators can be obtained through different ways: DSE~\cite{rw94} and numerical simulations in lattice field theory~\cite{hjr92}, for example.  

The paper is organized as follows: first, we present the model to calculate the $\gamma \gamma \rightarrow J/\Psi J/\Psi$ process using the approach of the ref.~\cite{km98}. Next, the main features of the employed non-perturbative gluon propagators are reviewed and then, we perform the calculation of the differential cross section, $d\sigma/dt$ in the two gluon approximation with a modified gluon propagator, and finally present the conclusions and discuss some possible future developments.


The differential cross section for the $\gamma\gamma \rightarrow J/\Psi J/\Psi$ obtained in the BFKL approach, is given by \cite{km98}
\begin{equation}	
    \frac{d\sigma}{dt} =  \frac{1}{16\pi}|{\cal A}(s,t)|^2 \label{eq1} 
\end{equation}
where
\begin{equation}
\Im m \, {\cal A}(s,t)  =  \int\frac{d^2 \k}{\pi} \,\frac{\Phi_0(k^2,Q^2)\,\Phi(x, \k, \Q)}{\left[ (\k + \Q/2)^2 + s_0 \right]\left[(\k -\Q/2)^2+s_0\right]}  \label{eq2}
\end{equation}
and $\k \pm \Q/2$ is the transverse momentum of the exchanged gluons; $s=W^2$ is the total center of mass energy of the $\gamma\gamma$ system; $t=-Q^2$ is the square of the transverse part of the momentum transfer (see figure \ref{f1}); $x=m^2_{J/\Psi}/W^2$, with $m_{J/\Psi}$ the $J/\Psi$ mass and $s_0$ is a parameter related to the employed gluon propagator (we will discuss about this point below). 

\begin{figure}[ht]
\begin{center}
\scalebox{.75}{\includegraphics*[75pt,420pt][512pt,760pt]{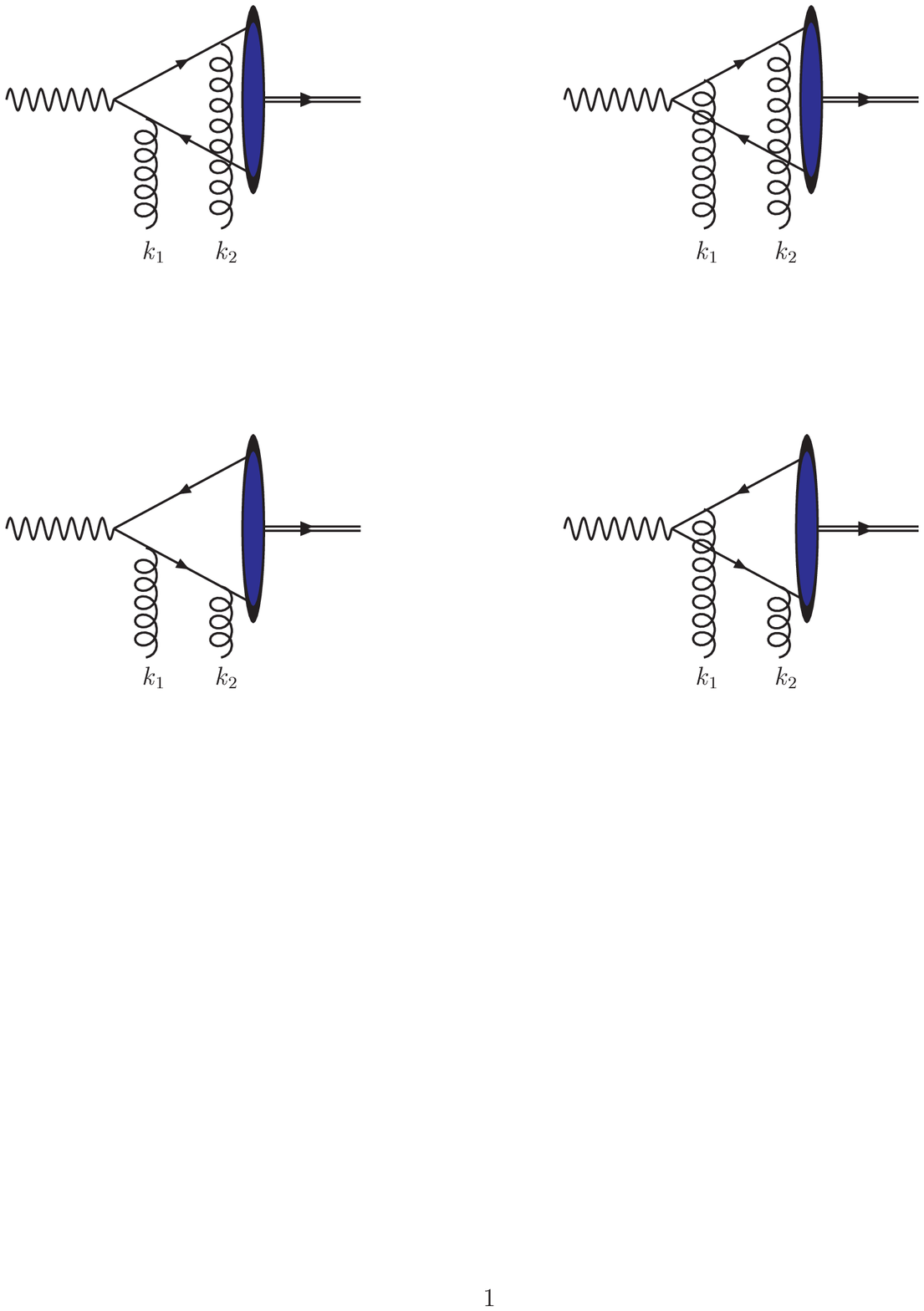}}
\end{center}
\caption{~The contributing Feynman diagrams for the form factor $\Phi_0$}
\label{f2}
\end{figure}

The impact factor for the transition $\gamma\,J/\Psi$ is $\Phi_0(k^2,Q^2)$ which is induced by two gluons (see figure \ref{f2}) and in a non-relativistic approximation~\cite{fz95} is
\begin{equation}
\Phi_0(k^2,Q^2) ={{\cal C}\over 2}\sqrt{\alpha_{em}}\alpha_s(\mu^2) \left[{1\over \bar q^2} - {1\over m_{J/\Psi}^2/4+k^2}\right]. \label{eq3}
\end{equation}
In the above formula,
\begin{equation}
{\cal C} = q_c{8\over 3} \pi m_{J/\Psi} f_{J/\Psi}, \label{eq4}
\end{equation}
where $q_C = 2/3$ is the charm quark charge; $f_{J/\Psi}$ is a parameter that characterizes the light-cone wave function of the charmonium and considered here with value 0.38 GeV~\cite{km98}; $\alpha_{em}$ is the electromagnetic coupling constant; $\alpha_S(\mu^2) = 12 \pi/(25\ln(\mu^2/\Lambda^2))$ is the strong running coupling constant in the one loop approximation (the QCD scale is $\Lambda = 0.23$ GeV), $\bar q^2 = (m_{J/\Psi}^2+Q^2)/4$ and $\mu^2 = k^2 + Q^2/4 + (m_{J/\Psi}/2)^2$.

The function $\Phi(x,\k,\Q^2)$ obeys the BFKL equation in the leading $\ln(1/x)$ approximation~\cite{bfkl}
\begin{eqnarray}
\Phi(x,\k, \Q) &=& \Phi_0(k^2,Q^2) + \frac{3 \alpha(\mu^2)}{2 \pi^2} \int^{1}_{x} \frac{dx'}{x'} \int d^2\k' \, \frac{1}{\k_0^2+s_0} \times \nonumber \\  
& &\left[R(\k_1, \k_2, \Q)\Phi(x',\k',\Q) - V(\k_1,\k_2,\Q)\Phi(x',\k,\Q) \right] \label{bfkl}
\end{eqnarray}
where $\k_0 = \k'-\k$, $\k_{1,2} = \Q/2 \pm \k$, $\k_{1,2}'=\Q/2 \pm \k'$ and 
\begin{eqnarray}
R(\k_1,\k_2,\Q) &=& \frac{\k_1^2}{{\k'}^2_{1}+s_0} + \frac{\k_2^2}{{\k'}^2_{2}+s_0} - Q^2\frac{\k_0^2+s_0}{({\k'}^2_1+s_0)({\k'}^2_2+s_0)},  \\
V(\k_1,\k_2,\Q) &=& \frac{\k_1^2}{{\k'}^2_1+{\k}_0^2+2s_0} + \frac{\k_2^2}{{\k'}^2_2+{\k}^2_0+2s_0}. 
\end{eqnarray}

The following gluon propagator is used in ref.~\cite{km98}:
\begin{equation}
\D(\k) = \frac{1}{\k^2+s_0}, \label{kmgp}
\end{equation}
based on the original paper of Balitski\v{\i} and Lipatov~\cite{bfkl}, where the parameter $s_0$ (which plays the role of a gluon mass) is generated using the Higgs mechanism. In \cite{km98}, the parameter $s_0$ is used to measure the magnitude of the infrared contributions to the cross sections. To scan these contributions, we will consider the following procedure: instead of the full BFKL Pomeron, we take the two gluon exchange and replace the perturbative gluon propagator by a non-perturbative one which includes the infrared contributions.

The choice of the non-perturbative gluon propagators can be justified by the fact that they include infrared effects and then, they can be used in a more natural way to describe processes in which the infrared contributions play an important role.  

Before the use of the non-perturbative propagators, we will resume the main features of the methods to calculate the gluon propagator and discuss its distinct forms.


The research of the infrared safe form of the gluon propagator has a long history~\cite{jem99}. The main trouble with the usual perturbative propagator is its divergence in the infrared region, more specifically, its pole at $\k^2=0$. Them, obtain an infrared safe functional form for the gluon propagator several methods, both numerical and analytical are used: Dyson-Schwinger equations, lattice field theory and renormalization group methods. These frameworks give a number of solutions with different behaviors in the infrared and ultraviolet regions, which depend on the gauge considered and the approximations selected, and that we will discuss separately in the following. 


The Dyson-Schwinger equations (DSE)~\cite{rw94} are an infinite tower of coupled integral equations among the Green functions of a field theory. For example, in QCD, the DSE for the quark self-energy involves the DSE for the quark and gluon propagators and the DSE for the quark vertex, and so on. In order to solve this system of equations, we must truncate it at some number of external legs and make an {\it ansatz} for the omitted part of the equations (normally using a Slavnov-Taylor identity). For example, in the case of DSE for the gluon propagator, we can restrict the number of external lines in the Green functions to three lines (no four gluon vertices), refuse the ghost propagator contributions and use the usual quark-gluon and three gluon vertices, resulting in a system with simplified equations for the gluon and the quark propagators. The solutions found in the literature (for the gluonic case) are very sensitive to the choice of the {\it ansatz}, to the choice of the gauge (Landau, Feynman, axial) and to the method used to solve the reduced set of equations, giving different behaviors in the infrared region: infinite, finite (zero and non-zero). Some examples of the gluon propagator obtained with this technique are summarized below.
\begin{itemize} 
\item Cornwall~\cite{corn82} used a gauge independent resummation of Feynman diagrams (known as pinch technique) to obtain a special set of DSE's, whose solution presents an effective gluon mass. In \cite{corn82} its value was determined as $m^{\rm eff}_g=500\pm200\,{\rm MeV}$. The main feature of this solution is a dynamically generated mass term in the propagator. Another remarkable feature of this solution is the correct ultraviolet behavior, according to the renormalization group.

\item  H\"abel {\it et al.}~\cite{hat90} used an alternative approach to calculate the Green functions for QCD. In the perturbation theory, the DSE's are solved by iteration, generating a power series for the propagators and vertices. In \cite{hat90}, the principle of perturbation theory is maintained but some {\it ansatz} in the functional form of non-perturbative expression of the Green functions are made, which results in a simplified set of DSE's. This set yields the following solution for the gluon propagator: 
\begin{equation}
\D_H(\k^2)=\frac{\k^2}{\k^4+b^4} \label{hgp}
\end{equation}
where $b$ is a parameter to be determined. Note that the propagator vanishes at $\k^2=0$ and is finite when $\k^2\rightarrow\infty$. Similar behavior was found by Gribov and Zwanziger~\cite{gz} using another arguments.

\item Gorbar and Natale~\cite{gn00} calculated the QCD vacuum energy as a function of the dynamical quark and gluon propagators using the effective potential for composite operators and using the operator product expansion (OPE) to relate the gluon and quark propagator with respectives condensates. The OPE gives the high energy behavior of the gluon polarization tensor $\Pi$, which is related with the gluon propagator through (in the Landau gauge)
\begin{equation}
\D^{\mu\nu}(\k^2)=-\frac{i}{\k^2-\Pi(\k^2)}\left(g^{\mu\nu}-\frac{k^\mu k^\nu}{\k^2}\right). \label{gnp}
\end{equation}
In \cite{gn00} the authors use the following {\it ansatz} for the gluon tensor polarization (the momentum is in the Euclidean space),
\begin{equation}
\Pi(K^2) =  \mu_g^2\,\theta(\chi \mu_g^2 - K^2)+\frac{\mu_g^4}{K^2}\,\theta(K^2-\chi \mu_g^2), 
\end{equation}
where
\begin{equation}
\mu^2_g  =  \left(\frac{34N\pi^2}{9(N^2-1)}\left\langle\frac{\alpha_s}{\pi}G^{\mu\nu}G_{\mu\nu}\right\rangle \right)^{\frac{1}{2}}, 
\end{equation}
and $N$ is the number of colors ($N=3$, in this case), $\langle(\alpha_s/\pi)G^{\mu\nu}G_{\mu\nu}\rangle\simeq\,0.01\,{\rm GeV}^4$ is the gluon condensate~\cite{svz79}, resulting in $\mu_g = \,0.61149\,{\rm GeV}$. The variational parameter $\chi$ is determined numerically in~\cite{gn00} as $\chi=0.966797$. 

\end{itemize}


We are aware of the use of the Lattice Field Theory (LFT) (for instance, see \cite{hjr92}) as a non-perturbative method based on first principles, to make predictions about the infrared sector of QCD. However, LFT is not completely free of approximations. A lattice is characterized by two parameters: its size (number of points) and the separation between the points, and, due to the computational limitations, it is impossible to simulate an infinite lattice with an infinitesimal separation. Then a finite (and small) lattice with a given separation is used, introducing the called finite size and separation effects. Additional problems are the fermion duplication, solved by the quenched approximation and the phenomena of Gribov copies. Similarly to the DSE approach, the final result is also gauge dependent. In a recent paper, Alexandrou {\it et al.}~\cite{aff01} studied the gluon propagator with the lattice method in the Laplacian gauge -- which is free of Gribov copies -- and found that the best fit to the numerical results is obtained with the Cornwall's propagator.


Having presented the features of the non-perturbative gluon propagators, we will employ them in a modified model to describe the process $\gamma \gamma \rightarrow J/\Psi J/\Psi$ in the infrared region, using the propagators obtained using the DSE approach, specially the Gorbar-Natale one~\cite{gn00} and the H\"abel one~\cite{hat90}.


As seen above, there are several different forms for the gluon propagator in the infrared region, nevertheless, these propagators can be tested in phenomenological applications. 

The most current model to describe the diffractive processes through non-perturbative gluons is the Landshoff-Nachtmann one~\cite{ln87}. In this model, the Pomeron consists of two non-perturbative gluons, whose usual properties are modified by QCD vacuum effects. This description of the Pomeron is very simple, since we don't include the multi-gluon exchange effects, described by the BFKL equation~\cite{bfkl}. A consequence of this simplicity is the independence of the total cross-sections on the energy. Another problem is the fact that the model is based on an Abelian gauge theory, contrary to  the non-abelian character of the QCD. Notwithstanding this, the LN model for the Pomeron, in connection with the Cornwall's propagator, was used to describe the elastic $pp$ scattering~\cite{hkn93} and elastic production of vector mesons~\cite{gdhn93} with good agreement with experimental results. 

Here we use the model of two non-perturbative gluons for the Pomeron for the process $\gamma \gamma \rightarrow J/\Psi J/\Psi$, and we consider the following approximations: first, we employ the Eq.(\ref{kmgp}) to identify the gluon propagator in the amplitude (Eq. (\ref{eq2})), and then we approximate the Pomeron as two non-perturbative gluons. With these assumptions, the amplitude is given by:
\begin{equation}
(\ref{eq2}) \Rightarrow \Im{\rm m} \, {\cal A}^{(0)}(s,t)  =  \int\frac{d^2{\bf k}}{\pi} \,\left[\Phi_0(k^2,Q^2)\right]^2 \D\left[({\bf k}-{\bf Q}/2)^2\right]\D\left[({\bf k}+{\bf Q}/2)^2\right] \label{amzero}
\end{equation}  
where $\D(\k^2)$ is the non-perturbative propagator.

\begin{figure}
\centerline{\psfig{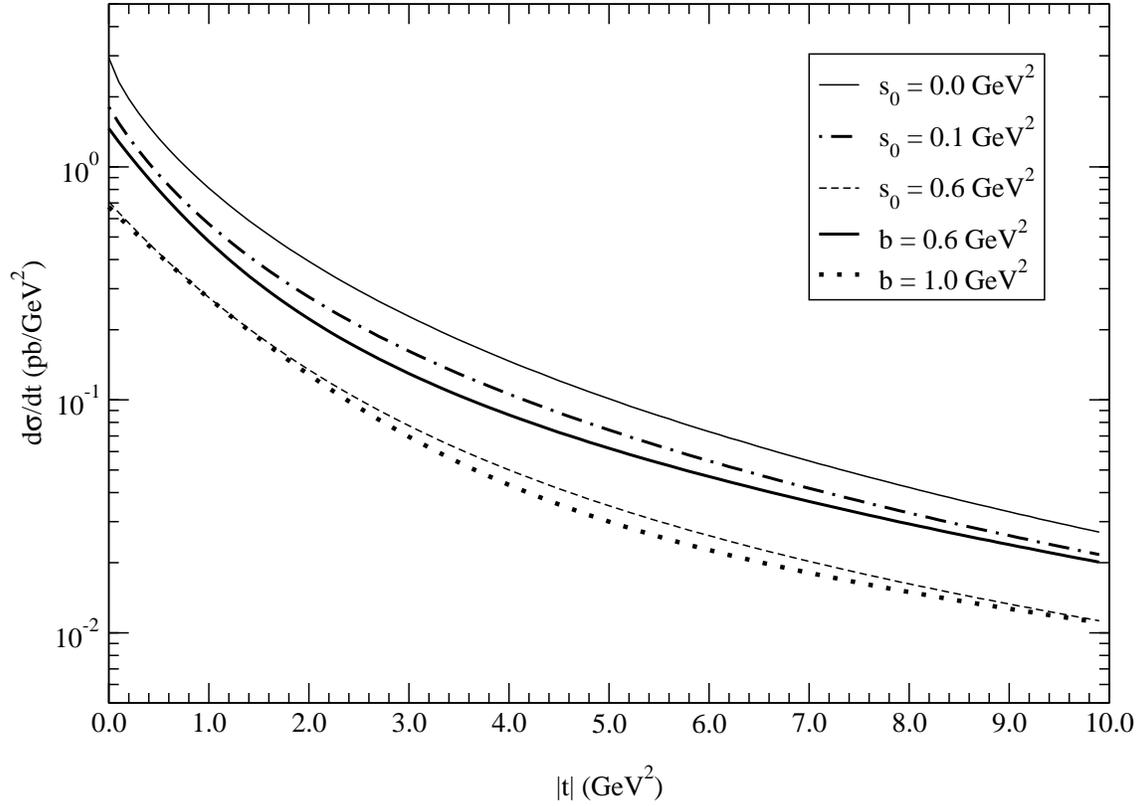}} \vspace{1cm}
\caption{The differential cross-section of the process $\gamma \gamma \rightarrow J/\Psi\,J/\Psi$ using the gluon propagators of Lipatov {\it et al.} (Eq.(\protect \ref{kmgp}), parameter $s_0$)~{\protect \cite{km98}} and H\"abel {\it et al.}(Eq.(\protect \ref{hgp}), parameter $b$).}
\label{f3}
\end{figure}

First, we substitute in the Eq.(\ref{amzero}) the H\"abel propagator (Eq.~\ref{hgp}) and vary the parameter $b$. This parameter is not determined in the works that proposed this propagator and we choose freely the values that give a result near to the one obtained in \cite{km98}.  Later, the results are compared with the propagator used in \cite{km98}~(Eq.~\ref{kmgp}), including the pure perturbative case (corresponding to $s_0=0$) as well as the pure massive case with a gluon mass similar to the one found by Cornwall~\cite{corn82}, $s_0=0.6\,{\rm GeV}^2$, since, in principle, we have freedom to set any value for it. As it can be seen in the results displayed in the fig.(\ref{f3}), the differential cross-section is sensitive to the choice of the propagator, but the global behavior is quite similar. The decrease of the cross section with both modified propagators is one order of magnitude or less, depending of the parameter involved and its value. When we compare the result of the perturbative case with the result of Eq.(\ref{kmgp}) with parameter $s_0=0.6\,{\rm GeV}^2$ in $|t|=5.0\,{\rm GeV}^2$, the cross-section decreases five times while for the H\"abel propagator with $b=0.6\,{\rm GeV}^2$ at the same $|t|$ value, the cross-section is two times smaller. The overall behavior is a significative decrease in the cross-section with both propagators. A remarkable feature is the similarity of the results from different propagators when $s_0=0.1\,{\rm GeV}^2$, $b=0.6\,{\rm GeV}^2$ and $s_0=0.6\,{\rm GeV}^2$, $b=1.0\,{\rm GeV}^2$. The result is also compatible with the fact that the H\"abel propagator tends to the perturbative one when the parameter $b$ goes to zero.

\begin{figure}
\centerline{\psfig{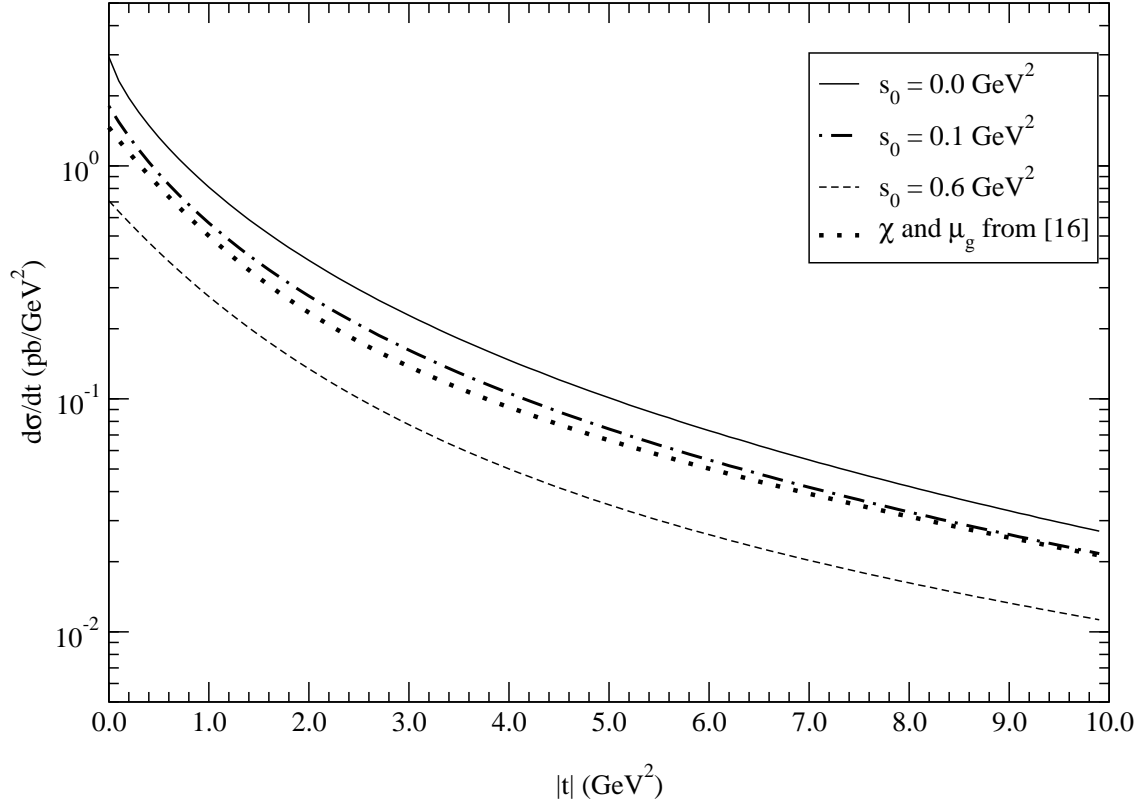}} \vspace{1cm}
\caption{The differential cross-section of the process $\gamma \gamma \rightarrow J/\Psi\,J/\Psi$ using the gluon propagators of Lipatov {\it et al.} (Eq.(\protect \ref{kmgp}), parameter $s_0$)~{\protect \cite{km98}} and Gorbar and Natale (Eq.(\protect \ref{gnp})) with the parameters obtained in {\protect \cite{gn00}}.}
\label{f4}
\end{figure}

In the next step we use the same strategy as above in the case of the Gorbar-Natale propagator, Eq.(\ref{gnp}), but we can not vary freely the parameters since that $\mu_g$ and $\chi$ depend on the value of the gluon condensate, experimentally obtained. For this reason, we use the values for $\mu_g$ and $\chi$ calculated in \cite{gn00}. The result is shown in the fig.(\ref{f4}) in comparison with the perturbative propagator, and is quite similar to the H\"abel propagator: with the choice of the parameters above the non-perturbative propagator gives a result near the perturbative one with a massive term with $s_0=0.1\,{\rm GeV}^2$. When this non-perturbative propagator is employed, the decrease of the cross-section in $|t|=\,0.5\,{\rm GeV}^2$  is around $40\,\%$. 

\begin{figure}
\centerline{\psfig{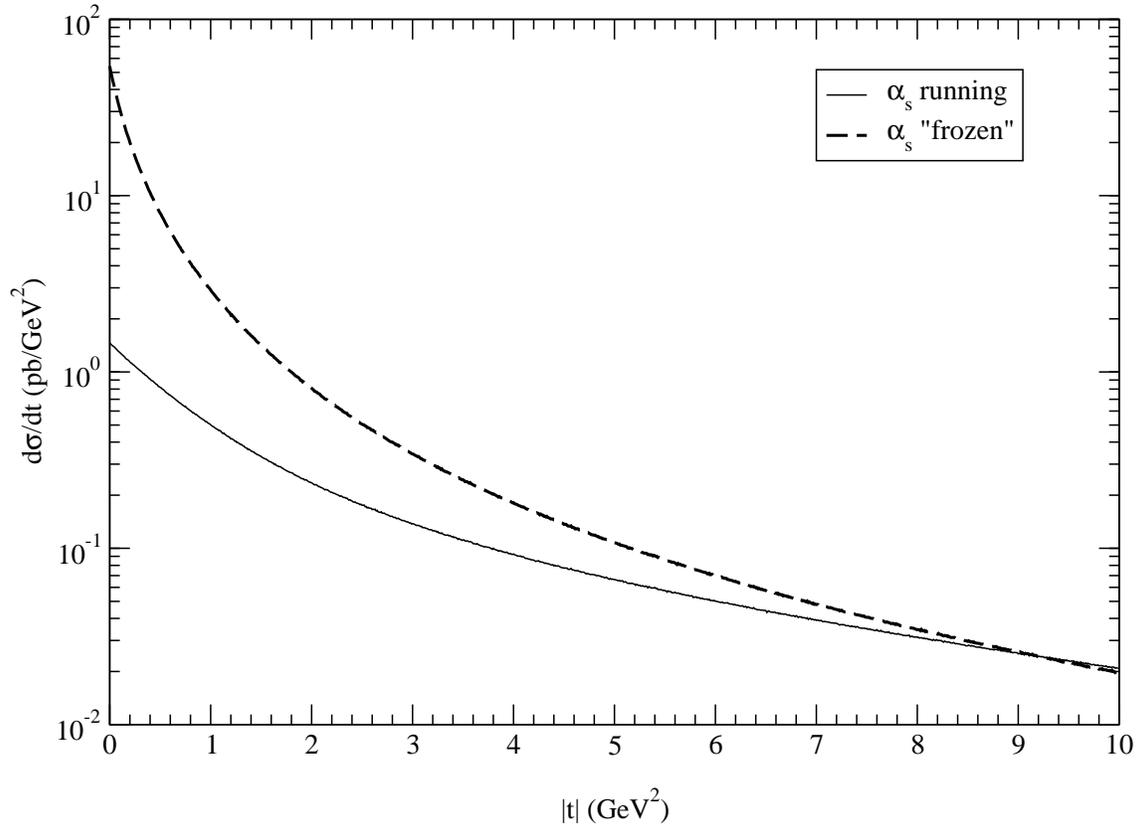}} \vspace{1cm}
\caption{Comparison between the results for differential cross-section obtained with a running coupling constant and frozen coupling with the Gorbar-Natale propagator~{\protect \cite{gn00}}. }
\label{f5}
\end{figure}

In the previous works in which the non-perturbative gluon propagators are used~\cite{hkn93,hpr96,gdhn93}, the running coupling constant is frozen at some value when $Q^2\rightarrow 0$. In the above results, we use a running coupling constant with a $k^2$ dependence when $Q^2$ goes to zero. In order to test the sensitivity of the cross-section with respect the choice of the coupling constant we computed the same observable, but now with a coupling constant with the above requirement, following the result from Cornwall~\cite{corn82}
\begin{equation}
\alpha_s^{C}(Q^2) = \frac{12\pi}{25 \ln(\frac{Q^2+4\mu_g^2}{\Lambda^2})} \label{asr}
\end{equation}
for the case of the Gorbar-Natale propagator. In the fig.(\ref{f5}) we shown the results, where the change of the coupling constant increases considerably the differential cross section (approximately two orders of magnitude) and modifies crucially the slope in the low values of $t$. 


The results obtained show that the behavior of the cross-section could be quite different (we obtain a decrease of one order of magnitude) once we change the gluon propagator. However, without experimental results expected in TESLA, we cannot restrict  reasonably the  choice of the propagator and the parameters for this process. The comparison with another diffractive process calculation obtained through the same methods could be employed, for example the elastic $pp$($\bar{p}p$) scattering or elastic photo-production of vector mesons, which we will soon consider.

Someone can argue against the use of the non-perturbative propagator in a perturbative calculation, but we only consider the first term in the perturbative expansion and the calculation of the form factor for the transition photon-vector meson is calculated in a non-perturbative manner, therefore our calculation is valid in the kinematical region of interest. More clearly, we are only interested in an estimation of the effects in the infrared region. We can see that with a particular choice of the parameters, the result with a bare gluon mass~\cite{km98} and the another one, with a non-perturbative propagator, is very similar (see figs.(\ref{f3}) and (\ref{f4})). It is well known that a massive gauge field theory like massive Yang-Mills presents several problems (see \cite{fpp99} for a review and use of this theory to describe diffractive phenomena) as well as the non-perturbative propagators but in the infrared region the last one seems to be a better choice than the former to describe diffractive processes. 

From our results, we can see that these effects in the region $|t| \leq 2\,{\rm GeV}^2$ diminish the differential cross-section for the distinct propagators once compared with the pure perturbative result. The goal of this work is to analyze the effects in the cross-section of the process $\gamma \gamma \rightarrow J/\Psi J/\Psi$ in the infrared region induced with the replacement of the usual gluon propagator by a distinct one, obtained from non-perturbative methods, which is more consistent in the kinematical region of interest. We show that the differential cross-section is reduced in comparison with the pure perturbative result in the very low momentum transfer regime with a running coupling constant. However, we should only be able to select which is the propagator that describes better the process once experimental results are provided. 

\newpage
\begin{center}
{\bf ACKNOWLEDGMENTS}
\end{center}

The authors thank A. A. Natale for clarifying discussions. W. K. S. thanks Magno V. T. Machado and L. Motyka for helpful suggestions in the early stages of the work. This work was partially financed by CNPq and by PRONEX (Programa de Apoio a N\'ucleos de Excel\^encia), Brazil.



\begin{references}

\bibitem{fr} J. R. Forshaw, D. A. Ross, {\sl Quantum Chromodynamics and the Pomeron}, Cambridge, 1997.

\bibitem{coll} Collins, P. D. B., {\it Introduction to Regge Theory and High Energy Physics}, Cambridge University Press, (1977).
 
\bibitem{bfkl} E.A. Kuraev, L.N.Lipatov and V.S. Fadin, {\sl Sov. Phys. JETP} {\bf 45} (1977) 199;  L.N. Lipatov, in "Perturbative QCD", edited by A.H. Mueller, (World Scientific, Singapore, 1989), p. 441 and references therein; Ya. Ya. Balitski\v{\i} and L. N. Lipatov. {\sl Sov. J. Nucl. Phys. }{\bf 28}, 822 (1978).

\bibitem{km98} J. Kwieci\'nski, L. Motyka, {\sl Phys. Lett. }{\bf B438}, 203 (1998).

\bibitem{der01} A. De Roeck, {\it Two Photon Physics at Future Linear Colliders}, {\sl hep-ph/0101075 } (2001).

\bibitem{hkn93} F. Halzen, G. Krein, A. A. Natale. {\sl Phys. Rev. }{\bf D47}, 295 (1993).

\bibitem{hpr96} D. S. Henty, C. Parrinello, D. G. Richards. {\sl Phys. Lett. }{\bf B369}, 130 (1996).

\bibitem{gdhn93} M. B. Gay Ducati, F. Halzen, A. A. Natale. {\sl Phys. Rev. }{\bf D48}, 2324 (1993).

\bibitem{rw94} C. D. Roberts, A. G. Williams. {\sl Prog. Part. Nucl. Phys. }{\bf 33}, 477 (1994).

\bibitem{hjr92} H. J. Rothe, {\sl Lattice Gauge Theory: An Introduction}, World Scientific, Singapure, 1992.

\bibitem{aff01} C. Alexandrou, Ph. de Forcrand, E. Follana. {\sl Phys. Rev.} {\bf D63}, 094504 (2001).

\bibitem{fz95} J. R. Forshaw, M. G. Ryskin, {\sl Z. Phys.} {\bf C68}, 137 (1995); I. F. Ginzburg, S. L. Panfil, V. G. Serbo, {\sl Nucl.Phys. }{\bf B296}, 569 (1988).

\bibitem{jem99} J. E. Mandula. {\sl Phys. Rep. }{\bf 315}, 273 (1999).

\bibitem{corn82} J. M. Cornwall, {\sl Phys. Rev. }{\bf D26}, 1453 (1982).

\bibitem{hat90} U. H\"abel, R. K\"onning, H.-G. Reusch, M. Stingl, S. Wigard. {\sl Z. Phys. }{\bf A336}, 423 (1990), {\sl Z. Phys. }{\bf A336}, 435 (1990).

\bibitem{gz} V. N. Gribov, {\sl Nucl. Phys.} {\bf B139}, 1 (1978); D. Zwanziger, {\it ibid.} {\bf B364}, 127 (1992), {\it ibid.} {\bf B378}, 525 (1992).

\bibitem{gn00} E. Gorbar, A. A. Natale, {\sl Phys. Rev. }{\bf D61}, 054012 (1994).

\bibitem{svz79} M. A. Shifman, A. I. Vainshtein, V. I. Zakharov, {\sl Nucl. Phys.} {\bf b147}, 385 (1979), {\it ibid.} {\bf B147}, 448 (1979). 

\bibitem{ln87} P. V. Landshoff, O. Nachtmann, {\sl Z. Phys. }{\bf C35}, 405 (1987).

\bibitem{fpp99} J. R. Forshaw, J. Papavassiliou, C. Parrinello, {\sl Phys. Rev. }{\bf D59}, 074008 (1999).

\end{references}
\end{document}